\documentclass[]{spie}  

\newcommand{\micron}{\textmu m}

\usepackage{amsmath,amsfonts,amssymb}
\usepackage{graphicx}
\usepackage{subcaption}
\usepackage[colorlinks=true, allcolors=blue]{hyperref}

\title{Spatial Mapping and Capacitor Trimming Developments to Improve Usable Pixel Yield in PRIMA FIRESS Kilo-Pixel Arrays}

\author[1,2]{Chris Albert}
\author[2]{Jonathon Hunacek}
\author[1]{Hien Nguyen}
\author[1,2]{Ritoban Basu Thakur}
\author[2]{Andrew Beyer}
\author[1,2]{Charles (Matt) Bradford}
\author[2]{Peter Day}
\author[2]{Pierre Echternach}
\author[2]{Byeong Ho Eom}
\author[1,2]{Logan Foote}
\author[2]{Marc Foote}
\author[1]{Steve Hailey-Dunsheath}
\author[1,2]{Reinier Janssen}
\author[1,2]{Elijah Kane}
\author[2]{Henry (Rick) LeDuc}
 
\affil[1]{California Institute of Technology, 1200 E California Blvd, Pasadena, CA 91125, United States of America}
\affil[2]{Jet Propulsion Laboratory, 4800 Oak Grove Dr, Pasadena, CA 91011, United States of America}

\authorinfo{Contact Author: Chris Albert (e-mail: calbert@caltech.edu)}

\begin{document} 
\maketitle

\begin{abstract}
The Probe far-Infrared Mission for Astrophysics (PRIMA) will use 8 kilo-pixel kinetic inductance detector (KID) arrays in its spectrometer module. We present an improved resonant frequency to spatial position mapping system designed to preserve each array’s mapping after transferring it from the mapping apparatus to the flight housing. Such a mapping is necessary for astronomical observations, and additionally allows us to laser trim the capacitive elements of KIDs to optimize resonance separation in frequency space. This increases the operating yield by eliminating collided resonances, reduces crosstalk, and reduces the sensitivity to frequency drift over time.
\end{abstract}

\keywords{Kinetic Inductance Detectors, Far-Infrared, PRIMA, Capacitor Trimming, Spatial Mapping}

\section{INTRODUCTION}
\label{sec:intro}  

The Probe far-Infrared Mission for Astrophysics (PRIMA) represents a major leap in far-infrared (far-IR) astronomy by improving sensitivity by 3 orders of magnitude and mapping speed by 6 orders of magnitude when compared to the previous generation of far-IR observatories~\cite{Glenn25}. This allows PRIMA to directly address the open astronomical questions outlined in the Astro2020 Decadal Survey~\cite{Astro2020} while also providing a platform for a wide range of guest-observer science~\cite{2023_PRIMA_GO}. An overview of the mission and its goals to study exoplanet formation and galactic evolution can be found in Glenn et al.~\cite{Glenn25}. PRIMA's performance improvement over previous far-IR missions is enabled by cryogenic 4.5~K optics, which contribute minimal noise from thermal emission, and ultrasensitive superconducting detectors. PRIMA will contain two instrument modules: PRIMAger~\cite{Laure25}, a hyper spectral imager and polarimeter, and the Far-IR Enhanced Survey Spectrometer (FIRESS)~\cite{Bradford25}, which will provide astronomical background limited observations at Sun-Earth L2 and cover a combined wavelength range of 24-261~\micron.

Both instruments are made with kinetic inductance detectors (KIDs)~\cite{zmuidzinas_2012}, which are superconducting LC resonators that respond to absorbed power. Power deposited in the inductive element of a KID will break Cooper pairs and increase kinetic inductance, leading to a measurable change in the resonant frequency. KIDs are highly frequency multiplexible, with many observatories~\cite{Meeker18, Adam18,Walter20, Wilson20, CCAT23} using $\sim$1000 detectors coupled to a single microwave feed line to efficiently package their detector arrays. Having this many detectors per feed line naturally presents the issue of matching each KID's physical location to its resonant frequency. Furthermore, deviation from designed resonant frequencies on large scale arrays will lead to a number of unmeasurable, collided resonances.

In this paper, we present the spatial mapping system developed for FIRESS's kilopixel arrays and the resonant frequency correction process for optimally spacing resonances. We iterate upon the mapping system from Albert et al.~\cite{Albert24} with significant hardware and software upgrades to create a system capable of rapid array processing and that preserves a flight-like electrical environment while mapping. We use a cryogenic light emitting diode (LED) array mapper design\cite{Liu17, Vav22, Shroyer22, Liu24} for its low cost and the minimal required cryostat modification. Unlike all previous LED-mapped arrays, which have KIDs sensitive to longer wavelengths than PRIMA, our FIRESS arrays are not designed with feedhorns that naturally collimate LED emission. FIRESS's targeted wavelengths (24-235~\micron) necessitate a lens-coupled design\cite{Cothard24, Dahal25}, as feedhorn fabrication becomes unfeasible at the short wavelength end of this range. The previous FIRESS collimated LED design~\cite{Albert24} produced sufficiently small spot sizes, but required placing the aluminum (Al) collimator within $\sim$100~\micron~of the detectors, causing electrical coupling not present in the flight-like measurement configuration. The new design solves this coupling issue by using a system of mirrors to reimage the collimated LEDs onto the KIDs, thus physically isolating the KIDs from any metal surfaces found exclusively in the mapping setup. The new, more sophisticated mapping pipeline has greatly increased analysis speed and improved mapping confidence.

Following the spatial mapping of a KID array, the resonances are precisely adjusted to maximize the frequency spacing between them by trimming the tines of the KIDs' interdigitated capacitors (IDCs). This increases the operating yield by separating collided resonances, while also improving the performance of the majority of the pixels by minimizing the component of electrical crosstalk that arises from resonator frequency spacing on the readout line. This maximal spacing for all KIDs also improves our overall margin against frequency drift or resonator confusion resulting from extended air exposure, in which each resonator slightly drifts lower in frequency at different rates as an oxide layer on the KID aluminum inductors thickens over time. Though other KID based instruments have achieved this by lithographically etching tines~\cite{Vav22}, we avoid re-processing the array in a cleanroom by using a high power laser to ablate the niobium capacitor tines. We demonstrate a high degree of frequency tuning control with this process and no adverse effects to detector performance from the destructive tine removal.

\section{Methods}
\label{sec: methods}

\subsection{LED Mapping}
\label{sec: mapping}

\begin{figure} [ht]
\begin{center}
\includegraphics[width=\textwidth]{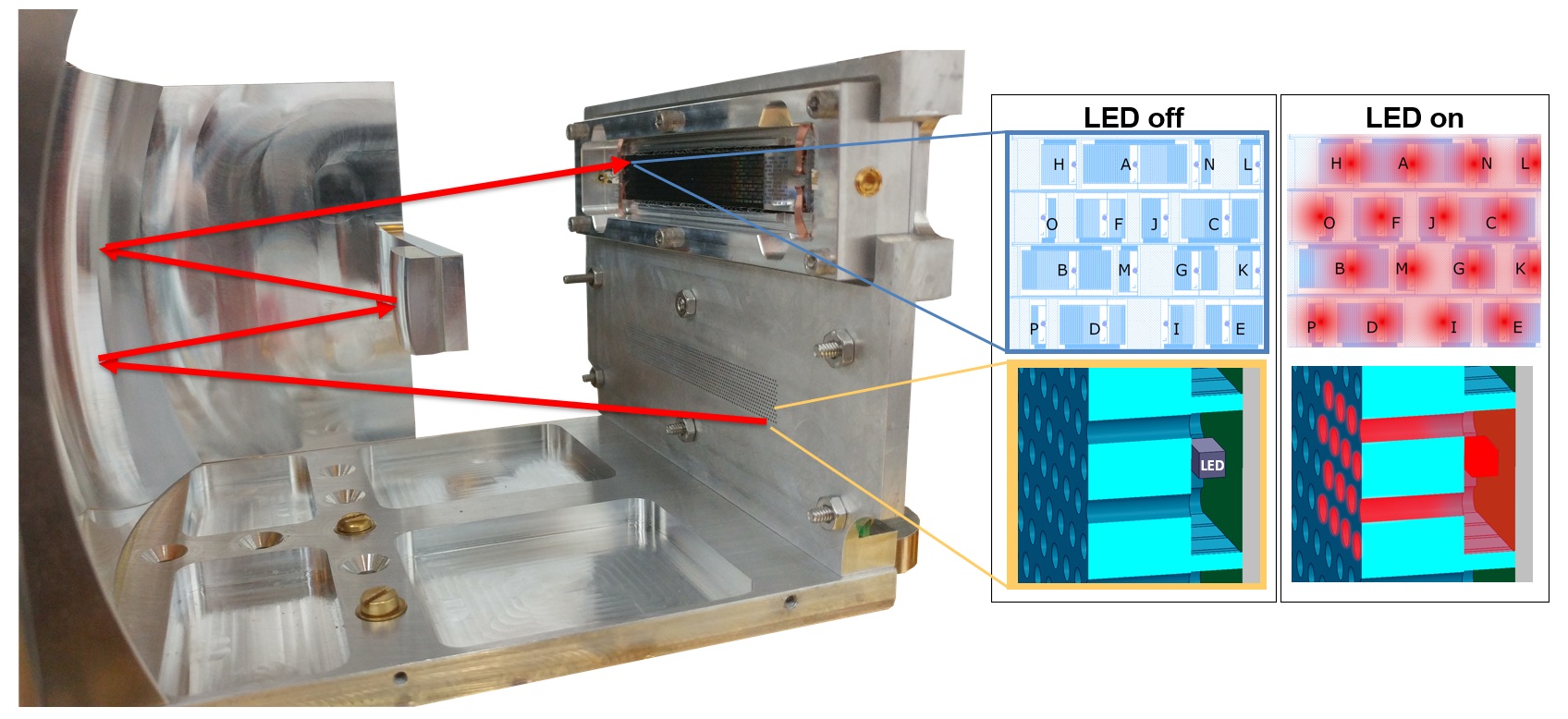}
\end{center}
\caption[] 
   {The reimaging mapping assembly with the illumination pattern of the collimated LEDs. The LED PCB is mounted on the backside of the bracket. When in operation, a blackened baffle extends from the light collimator up to the backside of the secondary mirror and a blackened sheet metal shell encloses the assembly for stray light isolation.}
\label{fig:map_setup}
\end{figure}

Each FIRESS chip contains 1008 KIDs capacitively coupled to a microwave feed line in a hex packed configuration with 900~\micron~spacing between absorbers. The array consists of repeated tilings of a 4$\times$4 KID unit cell. Within each unit cell, the resonances are logarithmically spaced in frequency by changing the number of tines in the IDCs. Across the array, the length of the tines in each unit cell is incremented, leading to 16 distinct banks of 63 resonances in frequency space, although in some cases the fabricated banks may overlap at the edges. The pixel arrangement inside a unit cell separates KIDs in neighboring banks from being spatial neighbors. Fig. \ref{fig:map_setup} shows the layout for current FIRESS short wavelength (24-43~\micron) design, with KID labels "A" through "P" corresponding to banks from lowest to highest frequency. With this design, one only needs to know the frequency bank and unit cell of a KID to determine its physical location.

To match KIDs to unit cells, we illuminate each unit cell using an array of 63 near-infrared LEDs (Kingbright APHHS1005F3C-70MAV). In principle, the strongest responding KID in each bank is assigned to the unit cell corresponding to that LED, though in practice, pixels lost during fabrication and light spillover complicate the analysis. To reduce spillover onto neighboring unit cells, the printed circuit board (PCB) containing the LED array is clamped to a collimator plate that isolates each LED and splits into 16 500~\micron~diameter holes that match the absorber arrangement of a unit cell. A thin gasket made of FR-4, the same material as the PCB, is placed between the collimator and PCB. It has a hole pattern that allows each LED to poke through, which strongly reduces LED spillover between neighboring collimator cells.

The collimated light is reimaged by an Offner relay onto the detector side of the KID array in a "flight-like" packaging designed with an electrical environment similar to the final flight packaging. Separating the collimator plate from the array housing is necessary to preserve this electrical environment and thus the resonant frequency distribution of all the KIDs on the array. The consistency of the resonant frequencies between the mapping system and standard operation is crucial to choosing the optimal capacitor trimming plan. The detector housing and mapping apparatus, shown in Fig. \ref{fig:map_setup}, are both made from an all aluminum construction so that alignment is preserved as the system is cooled to the KID operating temperature of 125~mK. Zemax OpticStudio was used to determine the detector and collimator placement. The collimator hole pattern is designed to contract to match the KID pattern after cooling down.

\begin{figure}[htbp]
    \centering
    \includegraphics[width=0.7\textwidth]{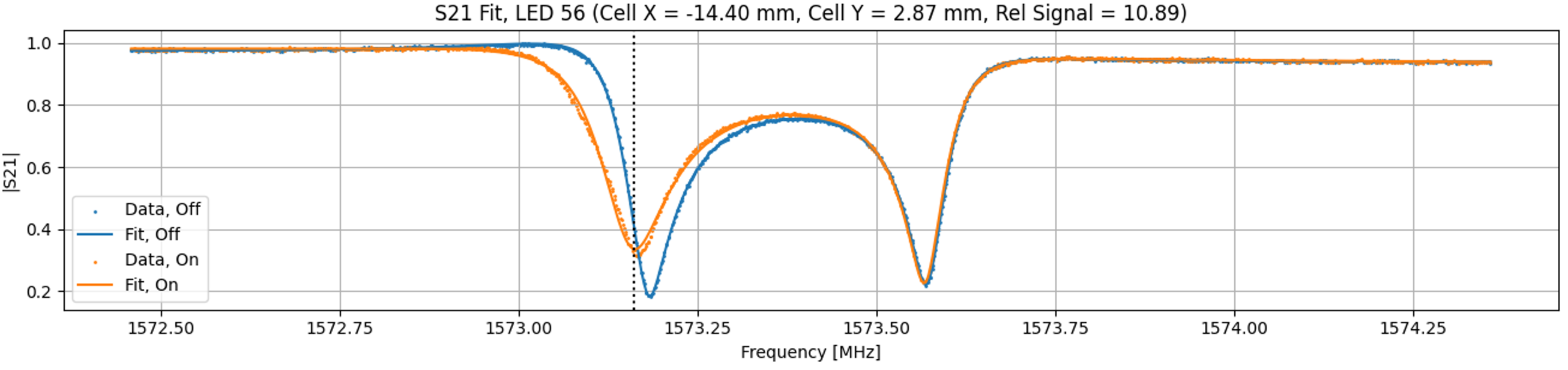}
    \caption{Magnitude of $S_{21}$ measurements and fits in one readout frequency segment for a single LED.}
    \label{fig:response}
\end{figure}

The data acquisition routine consists of measuring the frequency-dependent complex transmission ($S_{21}$) of all resonators while iterating through every LED. The LEDs are controlled with a room temperature relay board and biased  between 3-5~V, with a 40~M$\Omega$ resistor used to limit flux and prevent detector overloading. Complex $S_{21}$ sweeps around each resonator are taken repeatedly with one LED activated at a time, with reference measurements with no LEDs on recorded after each.  The $S_{21}$ sweep is broken into chunks of $n=$ 1 to 10 neighboring resonators, with each sweep for each chunk fit to the following approximation (which disregards feedline distance)

\begin{equation}
    S_{21} = z_0(f) \, \prod_{k=1}^{n} \left( 1 - \frac{(Q_{r,k}/Q_{c,k})}{\cos{\phi_k}} \, \frac{e^{i\,\phi_k}}{1 + 2i \, y_k} \right) 
    \;\;\;\;\;\;\;\;\;\;\;\;
    y_k - \frac{a_{nl,k}}{1 + 4 y_k^2} = \left( \frac{f - f_{0,k}}{f_{0,k}}  \right) Q_{r,k}
    \label{eq:fit}
\end{equation}

where $z_0(f)$ is a simple fitting model of the local complex feedline transmission, $Q_r$ is the total quality factor for each resonator, $Q_c$ is the coupling quality factor, $f_0$ is the resonance frequency, $a_{nl}$ captures nonlinear behavior, and $\phi$ captures impedance mismatch.  Sample fits for a single chunk with a single LED are shown in Fig. \ref{fig:response}.  The shift in resonant frequency and internal quality factor ($Q_i=1/(1/Q_r-1/Q_c)$) are extracted for each LED, forming a 3x21 coarse spatial map across the array.

\begin{figure}[htbp]
    \centering
    \includegraphics[width=0.7\textwidth]{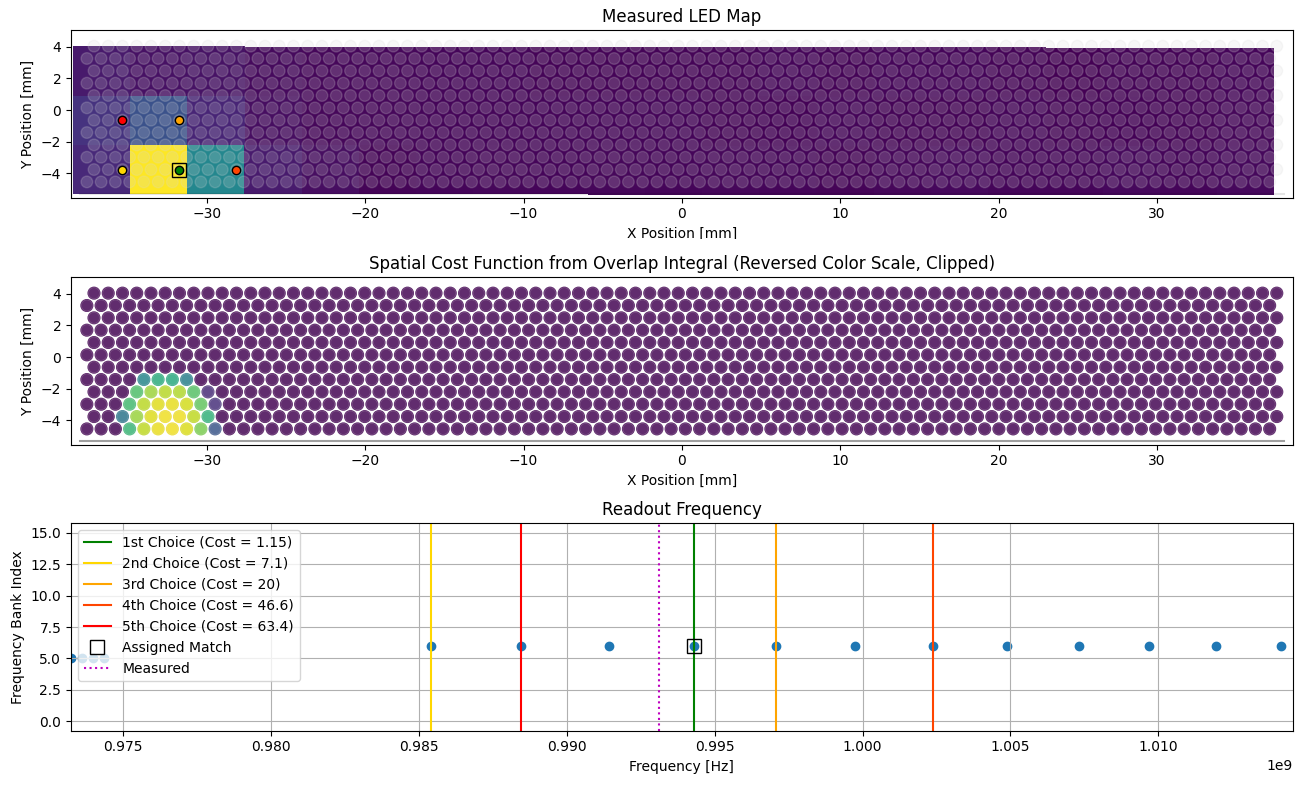}
    \caption{\textbf{Top:} Relative detector response (df/f) of a sample resonance to each LED. The top 5 position candidates are shown as colored points with the ultimately selected match indicated by a black square. \textbf{Center:} Spatial-only cost of each design pixel if it were to be matched with the measured resonance. \textbf{Bottom:} The predicted frequency of various pixels compared to the measured resonance, indicating the top candidates as colored vertical lines and final selection with a black square.}
    \label{fig:costs}
\end{figure}

A total cost matrix is calculated for each pairing of designed pixel position to observed resonance frequency across the entire array, which encodes our certainty that any given observed resonator corresponds to any specific physical pixel location.  The cost matrix is the sum of three terms: spatial, frequency, and quality. The spatial cost is determined by the overlap integral between the observed 3x21 LED map (e.g. top of Fig. \ref{fig:costs}) and the expected 3x21 map given the 16-hole-per-LED illumination pattern (plus a model of optics offset/alignment).  The overlap is computed for each possible pixel identity and defines a localized region on the array where the resonator resides (e.g. middle Fig. \ref{fig:costs}).  The frequency cost matrix is the distance between the observed resonator frequency and the expected frequency for every pixel, which provides the fine-scale information needed to break the degeneracy in pixel identification.  The quality cost matrix is computed from $S_{21}$ fit quality metrics and parameters, ensuring that poor fits do not dominate true matches.  From rows in the total cost matrix one can determine the most likely pixel position assignment for any given resonator, but this is not guaranteed to produce a one-to-one mapping.  Therefore, the total cost matrix is used as the metric for a stable matching algorithm~\cite{Gale62} which deterministically finds an optimal global matching solution. In Fig. \ref{fig:costs}, the five most likely matching candidates are shown in the top plot as colored points (with the measured LED response for that resonator as a background) and in the bottom plot as colored vertical lines, with the globally-optimal assignment indicated by a black square (matching the top choice in this, and most, instances).  Because array frequency schedule and optics alignment are not perfectly known, we then use the best-scoring matches to fit both a smooth readout frequency correction model and an optics alignment model (offset, rotation, scale).  The entire spatial and frequency cost functions are recomputed against these new baselines, and the stable matching algorithm is repeated.  This process repeats until jointly converging on a frequency model, an optics alignment model, and a list of pixel assignments.

\subsection{Laser Trimming}
\label{sec: trimming}

For each trimmed device, a frequency distribution algorithm determines the frequency shift required of each KID to maximize the separation of resonant frequencies within each bank and throughout the array. This process treats any remaining unidentified resonators as static obstacles to be avoided. The list of frequency shifts is then converted into a physical trimming plan. The amount to remove from a KID's IDC is given to first order by the simple linear sum of capacitor tine lengths, but is refined by physically-motivated correction terms which are functions of the per-pixel design geometries.  These higher-order terms are extracted from fits to the deviation between the desired trim and the measured result on previously trimmed FIRESS arrays. The refined trimming plan is consistent across devices, suggesting that the correction terms capture additional capacitive couplings in the design that are missed by an idealized IDC model. 

\begin{figure}[htbp]
    \centering
    \includegraphics[width=0.4\textwidth]{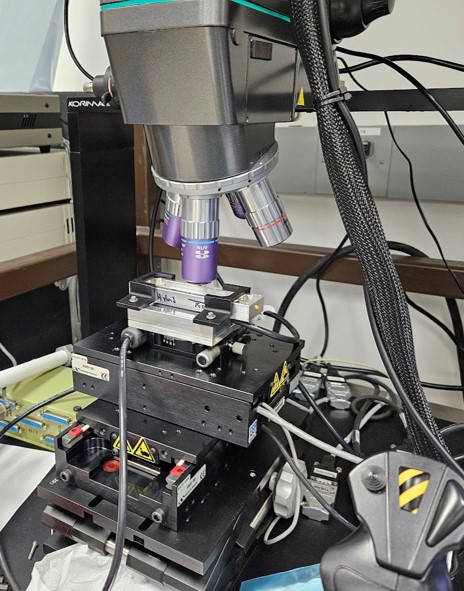}
    
    \vspace{0.2cm}
    \includegraphics[width=0.275\textwidth]{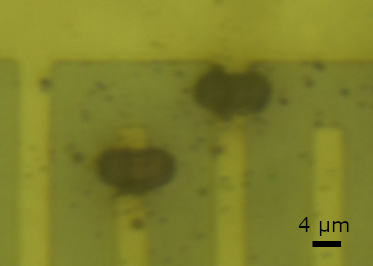}
    \includegraphics[width=0.28\textwidth]{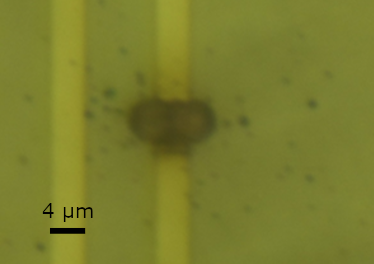}
    \caption{\textbf{Top:} Trimming setup with kilopixel array installed. The laser is contained inside the view-finder microscope. \textbf{Bottom:} Two examples of trimmed capacitor tines. Despite the localized debris spread by the laser ablation, no negative effects in performance were observed.}
    \label{fig:trim}
\end{figure}

The laser trimming apparatus is shown in Fig. \ref{fig:trim}. Like the mapping setup, this too is compatible with the FIRESS flight-like housing. This design allows detector arrays to be mapped, trimmed, and calibrated in an optical setup without ever needing to change housings. To control the trimming process, the KID housing is mounted to a programmable dual-axis linear translational stage. The trimming process is semi-automated, where custom stage control software automatically drives the laser's aim to each location specified by the device-specific trimming plan. An operator then focuses and fires the 532~nm laser while also verifying alignment with known marks at intervals along the stage's path of movement. 

\section{Results and Discussion}
\label{sec: results}

\begin{figure}[htbp]
    \centering
    \includegraphics[width=0.95\textwidth]{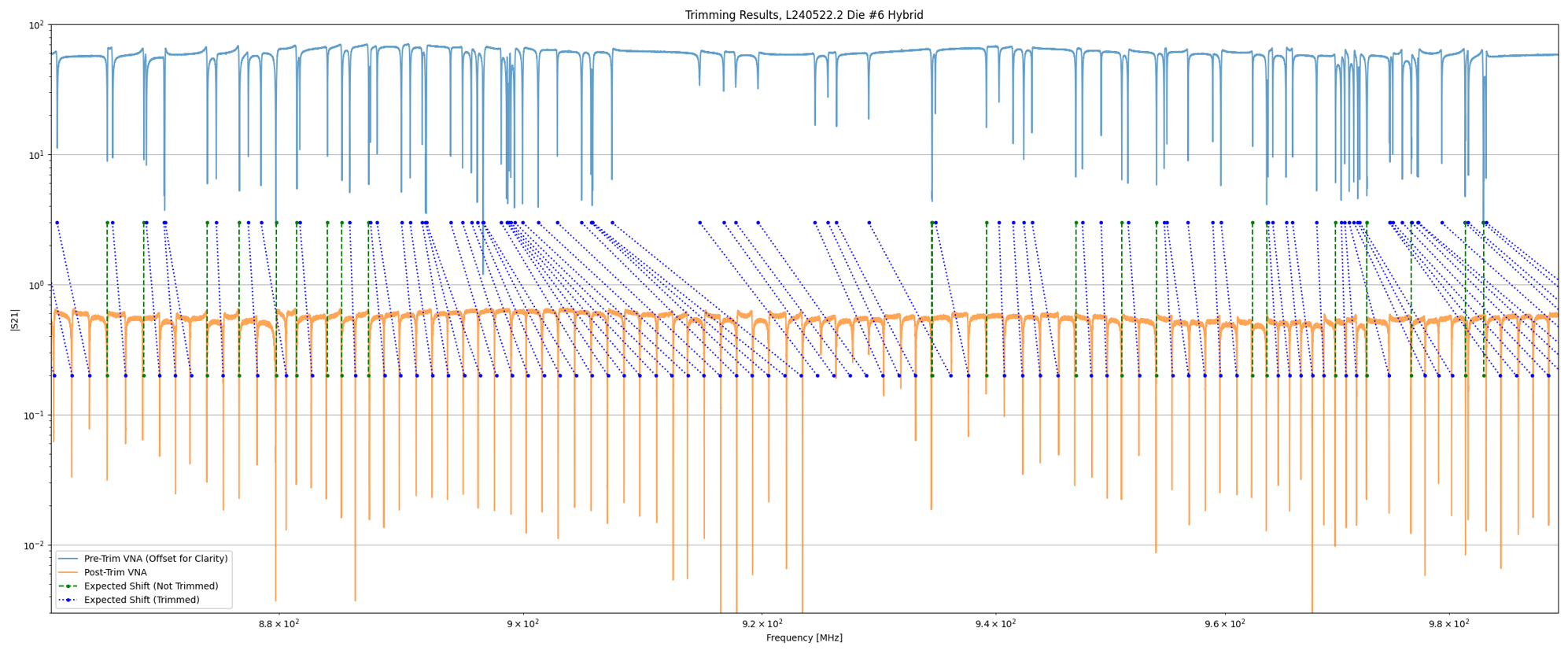}
        
    \vspace{0.3cm}
    \includegraphics[width=0.65\textwidth]{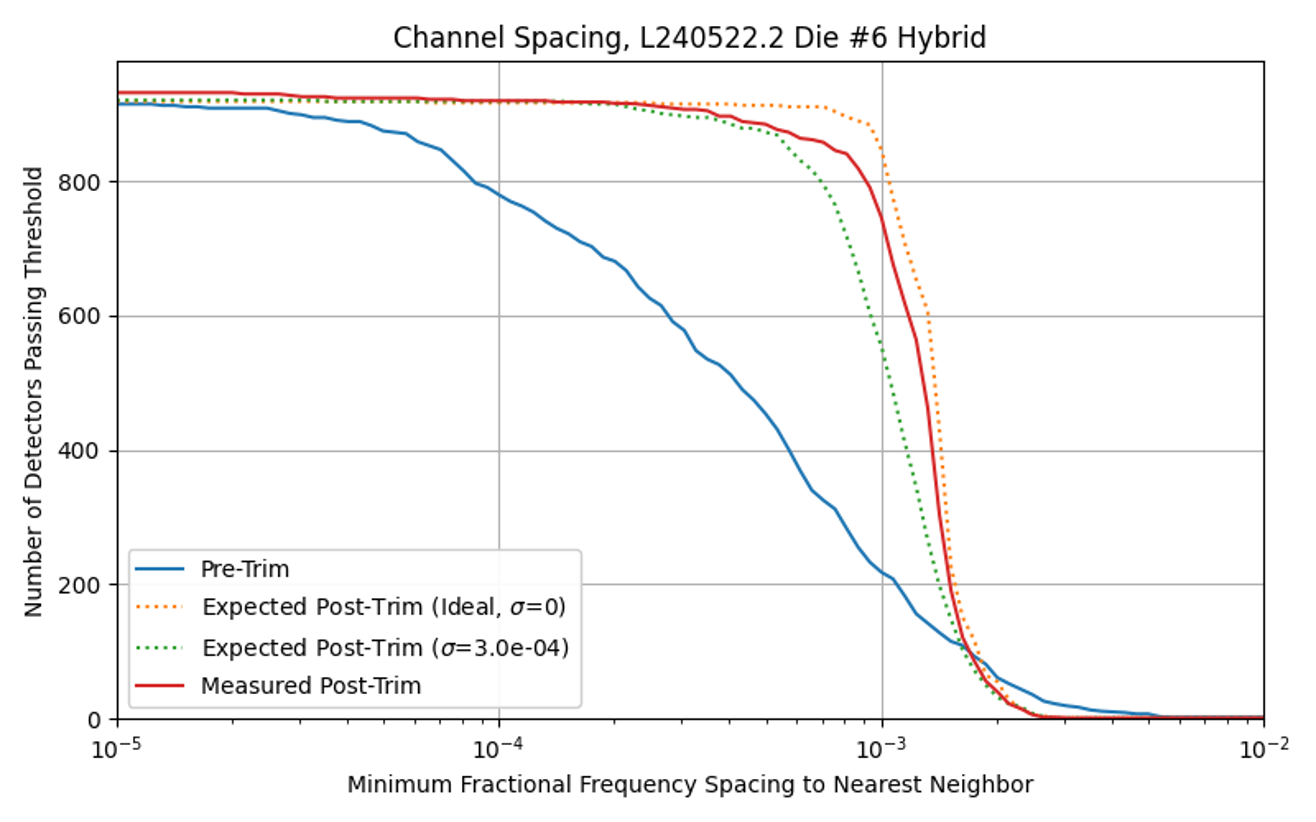}
    \caption{\textbf{Top:} The complex transmission magnitude, offset for visual clarity, of an array before and after trimming. The dashed lines indicate how much each resonator was shifted. The dashed blue lines are for trimmed resonators, while dashed green shows those not selected for trimming. \textbf{Bottom:} Fractional frequency spacing between resonances before and after trimming, with reference lines indicating the expected spacing distribution for different fractional frequency spacing standard deviations.  }
    \label{fig:corrected}
    
\end{figure}

We present trimming results on a microlens-hybridized 25~\micron-sensitive PRIMA kilopixel array. Out of 921 identified resonances assumed to be yielded KIDs, 914 were assigned matches, of which 876 are classified as high-confidence matches.  Of these, 773 were selected for modification by the frequency distribution algorithm, requiring 1023 total laser shots. The final frequency schedule generally matches the targeted distribution well, with the resulting fractional frequency spacing ($\delta x$) between resonances much more uniform. A sample of the pre- and post-trimming resonator spacings are shown in Fig. \ref{fig:corrected} (Top), where the original $S_{21}$ is shown in blue (offset for clarity), the trimming expectations are shown as dotted lines, and the measured post-trim $S_{21}$ is shown in orange. The expectation curves assume a Gaussian scatter in $\delta x$ about the target values, characterized by a standard deviation $\sigma$. Due to non-Gaussianity of the true distribution, we compare it to two bounding cases: $\sigma=0$ and $\sigma=3\times10^{-4}$. Over 90\% of the yielded KIDs in the post-trim array have $\delta x \geq 8\times 10^{-4}$. On the pre-trimmed array, the minimum distance to a nearest resonator for 90\% of the population was only $\delta x \geq 8\times 10^{-5}$.


\begin{figure}[htbp]
    \centering
    \includegraphics[width=0.8\textwidth]{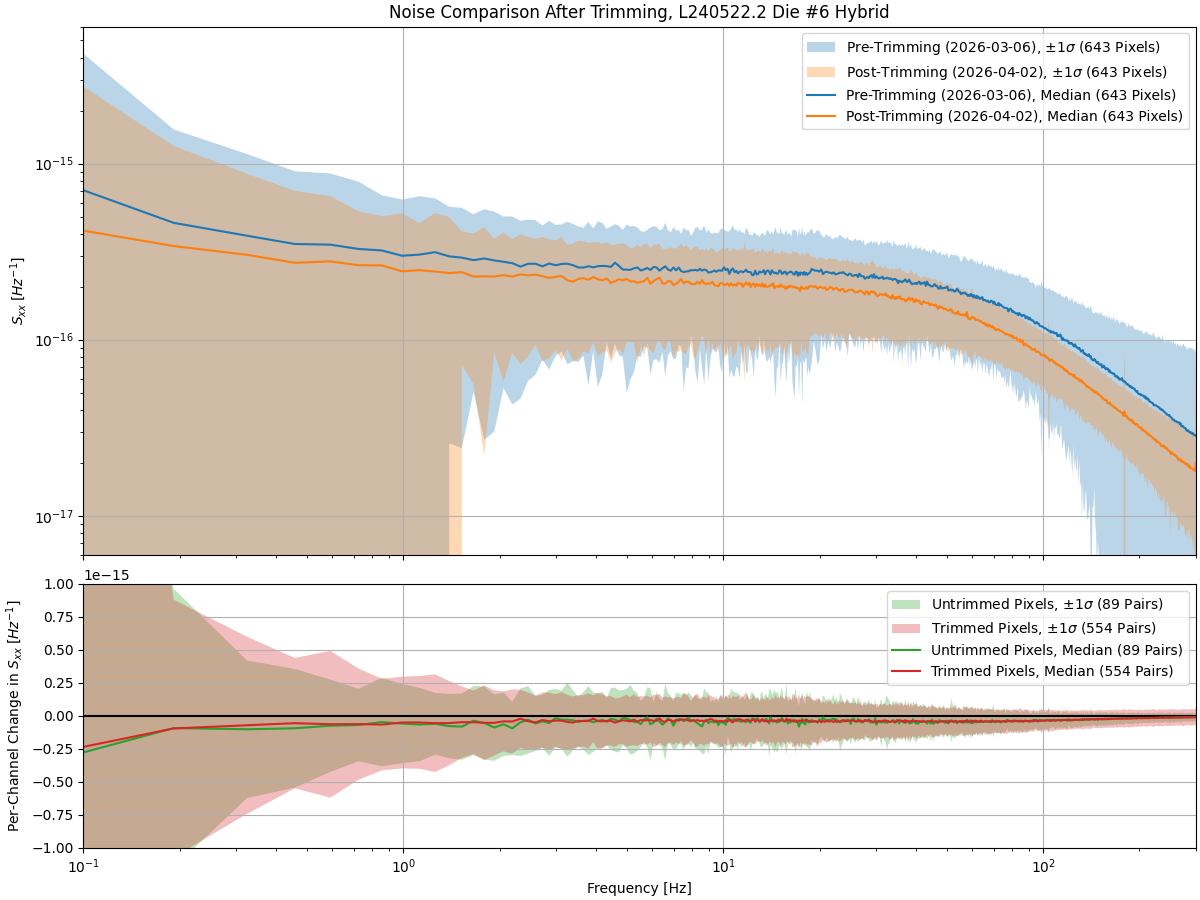}
    \caption{\textbf{Top:} Median $S_{xx}$ before and after trimming. \textbf{Bottom:} Median change in $S_{xx}$ for each pixel after trimming. There is no significant difference between the KIDs that had tines trimmed and those that did not, and both populations benefit from the array-wide reduction in crosstalk.}
    \label{fig:noise}
    
\end{figure}

After trimming, both the array's median detector noise and the variance of the noise across the array decreased. Any negative effects from the destructive trimming process were undetectable. The noise improvement is likely due to the reduction in crosstalk and the separation overlapping resonances. Fixing extreme cases of resonator collisions would remove outliers in the noise distribution, thus reducing the variance. Fig. \ref{fig:noise} shows the change in the fractional frequency noise power spectral density ($S_{xx}$) across the array and on a per-resonator basis.

\section{Conclusion}
\label{sec: conclusion}

The PRIMA FIRESS mapping and trimming process has reached a state of maturity capable of efficiently and accurately handling high throughput. A standard kilopixel array takes around 4 hours to measure in the LED mapper, with the automated analysis pipeline completed in a couple of hours on a standard PC (dominated by S21 fitting). The semi-automated trimming procedure requires 3–4 hours, significantly streamlining post-fabrication processing by avoiding cleanroom lithographic steps.

We are planning to upgrade the mapper by mounting the light collimator on a controllable, cryogenic dual-axis linear translational stage. This will allow for precise in-situ alignment. The current alignment is set by the tight mechanical tolerances of the components and a thermal contraction calculation. Minor misalignment can be identified by LED spillover from neighboring cells and is corrected in the analysis. As PRIMA approaches the production of flight candidate arrays, we anticipate using both the presented and planned mapper to double the mapping output per cryostat cooldown, as the cooldown schedule is a bottleneck in the processing speed.

\section*{Acknowledgments}

The research was carried out at the Jet Propulsion Laboratory, California Institute of Technology, under a contract with the National Aeronautics and Space Administration (80NM0018D0004). This work was funded by NASA (Award No.
141108.04.02.01.36) to Dr. C. M. Bradford. C. Albert was supported by the National Aeronautics and Space Administration (NASA) Space Technology Mission Directorate (STMD) through the NASA Space Technology Graduate Research Opportunities (NSTGRO) Fellowship under grant number 80NSSC24K1395 (PI: J. Zmuidzinas).

\bibliography{report} 
\bibliographystyle{spiebib} 

\end{document}